# Magnetic bistability and controllable reversal of asymmetric ferromagnetic nanorings

F. Q. Zhu,[1] G. W. Chern,[1] O. Tchernyshyov,[1] X. C. Zhu,[2] J. G. Zhu,[2] and C. L. Chien[1]*

[1]Department of Physics and Astronomy,
Johns Hopkins University, Baltimore, MD 21218

[2]Department of Electrical and Computer Engineering,
Carnegie-Mellon University, Pittsburgh, PA 15213

Magnetization reversals through the formation of vortex state and the rotation of onion state are two processes with comparable probabilities for symmetric magnetic nanorings with radius of about 50 nanometers. This magnetic bistability is the manifestation of the competition between the exchange energy and the magnetostatic energy in nanomagnets. The relative probability of the two processes in symmetric nanorings is dictated by the ring geometry and can not be altered. In this work, we report the magnetic behavior of a novel type of nanorings — asymmetric nanorings. By tuning the asymmetry we can control the fraction of the vortex formation process from about 40% to nearly 100% by utilizing the direction of the external magnetic field. The observed results have been accounted for by the dependence of the domain wall energy on the local cross section area of nanoring for which we have provided theoretical calculations.

*corresponding author: clc@pha.jhu.edu

The magnetic configurations and reversal mechanisms of nanomagnets depend intricately on their geometries and the competition between the magnetostatic and the exchange energies. For example, elongated nanomagnet can only acquire the single-domain state with associated magnetic poles at both ends and stray magnetic field in its vicinity. More interestingly, circular magnetic disc can acquire the vortex state in which the magnetization forms a closure that has no magnetic poles nor stray field [1-3]. This occurs in discs of sufficiently large radius $r$ where the magnetostatic energy prevails at the expense of the exchange energy. There is, however, a vortex core near the disc center that tends to disrupt the vortex state in discs with decreasing $r$. Indeed, when $r$ is below a threshold value $r_c$, which for Co and permalloy discs are in the range of one hundred to a few hundred nm [4], the vortex state cannot be accommodated. The bistability of single domain state and vortex state in Co nanodots has recently been reported [5].

A more intriguing geometry is that of a magnetic nanoring which has no central area and therefore contains no vortex core in the vortex state [6-12]. As a result, the vortex state can be stably retained in nanorings even for very small $r$. Ideal nanorings not only must have well-defined inner and outer radii, but also a narrow width less than that of a domain wall (about 50 nm in Co) to ensure that no vortex core can exist within the ring width. These dimensions for nanorings are challenging even for advanced electron-beam lithography which is commonly used for fabricating nanorings. The difficulty is further compounded by the need of a large number of nanorings for most measurements. Indeed, all reported studies of magnetic nanorings have been those with $r$ in the μm range, and few in the sub-μm range. Recently, some of us reported a new method using nanospheres as templates for the fabrication of large number ($10^9$) of magnetic nanorings with $r = 50$ nm and 20 nm in ring width [13]. These rings with $r = 50$ nm offer a new and hitherto unavailable medium for exploring the intricate magnetic properties of nanorings. On account of their vortex states, nanorings have been proposed for applications in high density magnetic recording and vertical magnetic random access memory (V-MRAM) [14].

The remnant state (at $H = 0$) of a magnetic nanoring after saturation is the so called "onion" state, consisting of two domains with semi-circular magnetizations of different helicity separated by two domain walls (DWs) on the opposite sides of the nanoring. Micro-magnetic simulation based on Landau-Lifshitz-Gilbert equation and experimental studies have revealed that under a magnetic field in the opposite direction, the onion state can switch via two different processes as shown in Fig.1 [13]. If the two DWs move towards each other in the beginning like in Fig. 1(a), they will be driven by the external field to move closer until annihilation to form the vortex state. If the two DWs move in opposite directions in the beginning, however, they will continue to do so until the onion state is fully reversed as illustrated in Fig. 1(b). These two processes are called the vortex formation process (V-process) and the onion rotation process (O-process) respectively. The probability $P$ of the V-process was found to be dependent on both the radius and the wall width of the ring [15]. For large ring with radius in the μm range, V-process is dominant. However, for radius $r$ in the 50 nm range, the two reversal processes have comparable probability. Indeed, in the case of $r = 50$ nm Co nanorings, we have found about 40% of them undergo the V-process while 60% through the O-process [13]. One notes that the bistability here is between the V-process and the O-process whereas in nanodots [5] it is between the single domain state and the vortex state.

It is important to understand the movement of DWs in nanorings to predictably achieve the vortex state in nanorings in the $r = 50$ nm size range, a feat also essential for applications employing magnetic nanorings. In this Letter, we report the special features of a new type of nanorings, which are *asymmetric* in its cross section. We show that the fraction $P$ of the V-process can be controlled by the direction of external field with respect to the axis of the asymmetric nanorings. In particular, when the magnetic field is applied along the asymmetry axis, the fraction $P$ is nearly 100%. We have also developed a theoretical model that can account for the preferred vortex formation reversal mechanism. The introduction of asymmetry in the nanorings allows full vortex formation without losing the virtue of small dimension, high stability and high areal density.

We used $r = 50$ nm polystyrene (PS) spheres as the templates in our fabrication process as shown in Fig. 2. A monolayer of isolated PS spheres were chemically attached to a clean Si(100) surface. A 40 nm thick Co film was sputter-deposited from a 99.995% pure Co target in a magnetron sputtering system with an *in situ* substrate tilting-angle adjusting capability. The base pressure and Ar sputtering pressure are $6 \times 10^{-8}$ Torr and 6 mTorr respectively. The substrates were swept across the sputtering plasma in order to have a uniform thickness everywhere. A broad beam $Ar^+$ ion source was then used to etch away all Co except those protected under the PS spheres resulting in Co nanorings. When the ion beam was at normal incidence to the substrate surface [$\alpha = 0$, Fig. 2(b)], one obtained nanorings of uniform width, or symmetric nanorings. However, when the substrate was tilted by an angle $\alpha$ as shown in Fig. 2(c), one obtained asymmetric nanorings in which the cross section at one side is larger than the other side. A capping layer of 5 nm Au was deposited for protection against oxidation. In the following, asymmetric nanorings fabricated with $\alpha = 10^o$ and $14^o$ are denoted as AR10 and AR14 respectively.

Fig. 2(d) and (e) are the top view scanning electron micrographs of the symmetric and asymmetric nanorings respectively. The cross section area in the symmetric nanoring is constant along the circumference, but varies in sample AR14, with the left side much wider and thicker than the right side. The asymmetry of AR14 is also revealed in the inset of Fig. 2(e) taken with a composition sensitive detector. The maximum and minimum ring widths of AR14 are about 60 nm and 10 nm, respectively.

The magnetic switching properties were measured with an ADE Model 10 vector vibrating sample magnetometer at room temperature. To reveal the effect of asymmetry on switching properties, measurements were made with the magnetic field applied in the substrate plane along different angle $\theta$ with respect to the symmetry axis (bottom right inset of Fig. 3, pointing from A to A'). Ten hysteresis loops of sample AR14 measured at $\theta = 0^o$ to $90^o$ are plotted together in Fig. 3, all showing the two-step switching characteristics but with a systematic variation as $\theta$ is varied. The highlighted area represents the field range in which the magnetization has substantial changes when the field angle varies. For comparison, the top left inset shows that the hysteresis loops of the symmetric nanorings are independent of $\theta$ as expected.

The measured hysteresis loops were fitted with the simulated switching loops of the V-process and the O-process as shown in Fig. 1 to determine the chance of acquiring the vortex state. The V-process shows a two-step reversal with two switching fields between which the ring retains the vortex state. The magnetization of the vortex state in a magnetic field is not exactly zero but increases weakly with the field until the second switching field is reached. This weak increase comes from the slight tilting of every atomic moment by the external field. The average magnetization of vortex state $M_{Vor}$ [indicated by the cross in Fig. 1(a)] between the two switching fields is therefore not zero but a small finite value. However, the O-process shows one-step reversal, and the average magnetization $M_{Rot}$ [indicated by the cross in Fig. 1(b)] at the same field is large and close to the saturation magnetization. When both processes are present as in a collection of symmetric nanorings, the resultant hysteresis loop is the superposition of these two loops weighted by the fractions $P$ and $1$-$P$ respectively. Consequently, the corresponding average magnetization value $M_{Mid}$ between the two switching fields will be $M_{mid} = P \times M_{Vor} + (1-P) \times M_{Rot}$. In Fig. 3, the location of $M_{Mid}$, indicated by the black line, depends systematically on the field direction. When the magnetic field direction is changed from the symmetry axis ($\theta = 0°$) to the asymmetry axis ($\theta = 90°$), $M_{Mid}$ decreases progressively and becomes nearly zero at $\theta = 90°$. The fraction $P$ can be deduced as $P = (M_{Rot} - M_{Mid})/(M_{Rot} - M_{Vor})$ from the above formula. The derived values of $P$ in symmetric nanorings and asymmetric nanorings of AR10 and AR14, are plotted together in Fig. 4 as a function of the field angle $\theta$. In symmetric nanorings, $P$ is about 38% and does not depend on the field direction. However, for asymmetric nanorings, the value of $P$ shows a strong dependence on the field direction, which changes from 39% to 73% in AR10, and from 41% to 98% in AR14. Therefore we not only have observed the magnetic bistability of the asymmetric nanorings, but also have achieved tuning of the bistability with the direction of the external field. The value of $P = 98\%$ at $\theta = 90°$ for AR14 shows that nearly every nanoring reverses its magnetization through the V-process, an important attribute for applications.

The fraction $P$ for the V-process is a direct measure of the magnetic bistability. In very large rings where the magnetostatic energy dominates, only V-process occurs ($P = 1$). In the other extreme of very small rings where exchange energy dominates, the magnetic reversal undergoes only the O-process ($P = 0$). For the intermediate sizes of a few hundred nanometers, both processes can occur with certain probabilities. The motion of DWs at the onset of the magnetization reversal determines the type of the reversal process, as shown by the insets in Fig. 1. For a symmetric nanoring with constant cross section, $P$ is intrinsic to its dimension and cannot be altered. In contrast, in asymmetric nanorings $P$ can be varied greatly by exploiting the field angle. In the following we address the effects of cross section area on the switching behavior of asymmetric nanorings.

The key aspect is to determine the dependence of the domain wall energy $E$ on the local cross section area of the nanoring. To this end we use a straight strip of constant width $w$ and thickness $t$ to approximate one segment of the ring and consider the energy $E$ of a domain wall inside. The general trend is $E$ grows with both $w$ and $t$. Even in the absence of intrinsic anisotropy in the material, the problem is computationally difficult because of the long-range magnetostatic interaction. Here we demonstrate this fact in a particular regime, when the ring is thin and narrow. The computation simplifies in the thin-film limit of $t \ll w$ and $tw \ll \lambda^2 \ll wt\log(w/t)$, where $\lambda = \sqrt{A/\mu_0 M_0^2}$ is the magnetic length (3.8 nm in Co) [16, 17]. In this geometry the magnetostatic term becomes local making an analytical solution possible. Owing to the shape anisotropy, the magnetization lies in the plane of the film. The magnetic energy is the sum of the exchange and magnetostatic terms:

$$E[\hat{\mathbf{m}}(r)] = \int_\Omega |\nabla\hat{\mathbf{m}}|^2 d^2r + (1/\Lambda)\int_{\partial\Omega} (\hat{\mathbf{m}}\cdot\hat{\mathbf{n}})^2 dr \qquad (1)$$

where $\hat{\mathbf{m}} = (\cos(\theta), \sin(\theta))$ is a unit vector depicting the in-plane components of the magnetization, $\hat{\mathbf{n}} = (0,1)$ is the normal to the edge, and $\Lambda = \lambda^2/t\log(w/t)$ is the thin-film magnetic length [16, 17]. $\Omega$ is the two-dimensional strip $-w/2 < y/w/2$, and $\partial\Omega$ is its boundary. The energy is made dimensionless by dividing out the unit of energy $\mu_0 M_0^2 \lambda^2 t = At$. Eq. (1) is in fact the familiar XY model with anisotropy at the edge added by the magnetostatic term. The ground states are uniform with $\theta = 0$ or $\pi$.

Minimization of the energy Eq. (1) yields the Laplace equation $\nabla^2\theta = 0$ and the boundary conditions $\partial_y\theta = \mp(1/\Lambda)\sin 2\theta$ at the upper and lower edges $y = \pm w/2$. A domain-wall solution interpolating between the ground states has the following structure:

$$\tan\theta(x,y) = \cos ky / \sinh k(x - X) \qquad (2)$$

where $X$ is the horizontal coordinate of the wall and the wavevector $k$ is determined by the boundary condition. In the thin-film limit, $k \approx \pi/(w + 2\Lambda)$. The energy Eq. (1) of the domain wall evaluates to $E \approx 2\pi(1 + \log(w/\pi\Lambda))$. As expected for the XY model, the exchange energy depends logarithmically on the strip width $w$ and on the short-distance cutoff $\Lambda$. After restoring the energy units and expressing $\Lambda$ in terms of the relevant parameters we obtain the final result for the energy of the domain wall

$$E \approx 2\pi At \log\left(\frac{ewt\log(w/t)}{\pi\lambda^2}\right) \qquad (3)$$

Note that, in the thin-film limit, where exchange energy dominates, the energy of the domain wall depends most sensitively (linearly) on the thickness $t$ of the strip, whereas the dependence on the width $w$ is rather weak (logarithmic). Our numerical evaluation and dimensional analysis [18] show that in thicker and wider rings, $E$ is still an increasing function of $w$ and $t$. We can apply Eq. (3) to estimate the domain wall energy inside the nanorings.

In asymmetric nanorings, the ring width and thickness vary along the circumference, with their minima and maxima separated by $180^{\circ}$. If the initial magnetic field is along the symmetry axis, one domain wall (DMW1) will be generated at the thinnest location A and the other domain wall (DMW2) at the thickest location A' after the field is removed (Fig. 3). Under a reversal magnetic field, DMW1 and DMW2 still have two possible directions to move, and the situation is not very different from that of the symmetric rings. The vortex probability $P$ is close to that of a uniform nanoring. However, if the initial field is along the asymmetry axis, DMW1 and DMW2 will be generated at the middle locations B and B' with the largest slope. Both domain walls have the tendency to slide to the same energy minimum position A' albeit the pinning forces of the local defects and roughness. As a consequence, the V-process is enhanced. The higher the asymmetry and the larger the geometrical slope, the higher $P$ will be.

In summary, we have observed the magnetic bistability between the V-process and the O-process of asymmetric nanorings fabricated with oblique angle ion beam etching. Comparing with the symmetric nanorings, asymmetric nanorings have controllable magnetic switching properties depending on the direction of the external field. We have achieved nearly 100% vortex reversal in the asymmetric nanorings while the symmetric nanorings can only accommodate 40%. We have also computed the energy of the domain wall, which depends on the local thickness linearly. The variation of local cross section in asymmetric nanoring favors the domain wall motion towards the thinnest position thus enhancing the vortex formation process.

**ACKNOWLEDGMENTS**

This work has been supported by NSF Grant No. DMR05-20491.

**Captions**

**FIG. 1.** (Color online) Micromagnetic simulation of (a) the vortex formation process and (b) the onion rotation process. Magnetizations $M_{Vor}$ and $M_{Rot}$ in the middle of the two switching fields are indicated by cross symbols. Insets illustrate the motion of domain walls at the onset of reversal.

**FIG. 2.** (Color online) Fabrication schematics of nanorings. (a) Si substrate attached with a layer of PS spheres is etched by $Ar^+$ ion beam at (b) normal angle for normal nanorings, and at (c) an oblique angle for asymmetric ones. A protective capping layer of Au is sputtered at the last step. Top view SEM micrographs of (d) symmetric and (e) asymmetric nanorings fabricated from $r = 50$ nm PS spheres at $0°$ and $14°$ ion milling angle. Inset: Composition sensitive SEM image of the ring, bright areas represent Co.

**FIG. 3.** (Color online) Evolution of hysteresis loops of $r = 50$ nm asymmetric nanorings at various magnetic field directions between $0°$ (inner most curve) and $90°$ (outer most curve). The highlighted area represents the field range in which the magnetization has substantial changes. Bottom right inset: The symmetry axis is defined as the direction pointing from the widest part A to the thinnest part A', the asymmetry axis is $90°$ away from it, from B to B'. Top left inset: Hysteresis loops of the uniform nanorings measured along various in-plane field directions are essentially the same.

**FIG. 4.** (Color online) Field angle dependence of the probability $P$ of the vortex formation process for symmetric nanorings (solid dots), AR10 (open squares) and AR14 (open circles). The connecting lines are the guide for eyes.

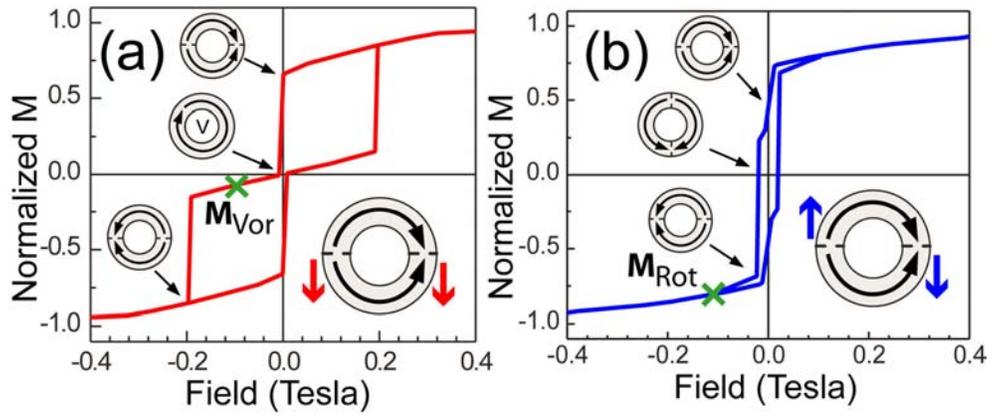

Fig. 2

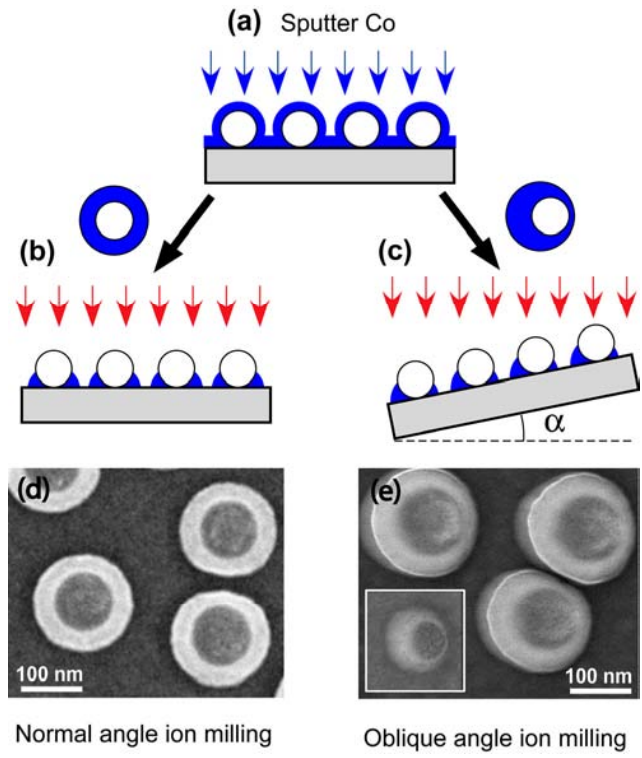

Normal angle ion milling        Oblique angle ion milling

Fig. 3

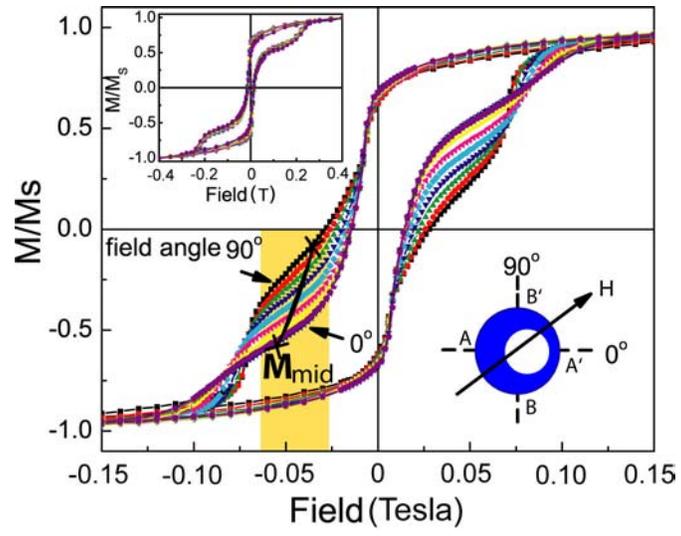

Fig. 4

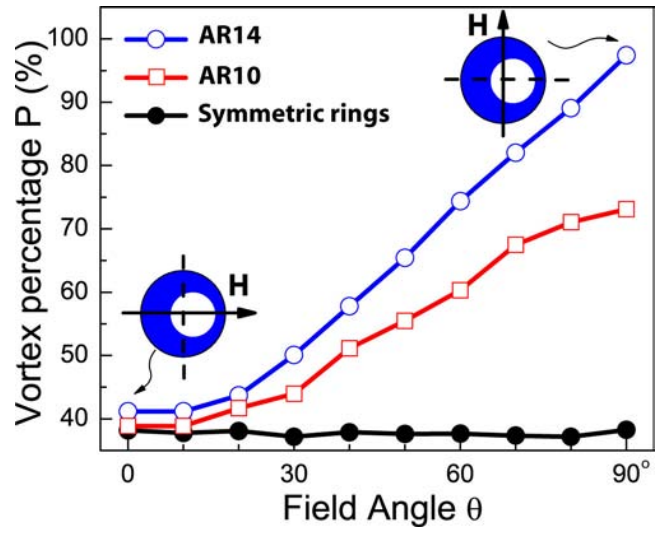